\documentclass[12pt,preprint]{emulateapj}

\usepackage{natbib}
\usepackage{color}

\shorttitle{High-speed plasma outflows in solar impulsive events}
\shortauthors{Wang, Sui, \& Qiu}

\begin{document}

\title{Direct observation of high-speed plasma outflows produced by magnetic reconnection
in solar impulsive events}

\author{TONGJIANG WANG\altaffilmark{1}, LINHUI SUI\altaffilmark{2,3}, JIONG  QIU\altaffilmark{1}\\
Received 2007 February 22; accepted 2007 April 20; published 2007 May 15\altaffilmark{4}
}
\altaffiltext{1}{Department of Physics, Montana State University, Bozeman, MT; wangtj@solar.physics.montana.edu}
\altaffiltext{2}{Department of Physics, Catholic University of America, 620 Michigan
    Avenue, Washington, DC}
\altaffiltext{3}{Heliophysics Science Division, Code 671, NASA Goddard Space Flight Center, Greenbelt, MD}
\altaffiltext{4}{\textcolor{blue}{Published in ApJ, 661, L207--L210, 2007}}

\begin{abstract}
Spectroscopic observations of a solar limb flare recorded by SUMER on SOHO reveal, for the first time, 
hot fast magnetic reconnection outflows in the corona. As the reconnection site rises across the SUMER 
spectrometer slit, significant blue- and red-shift signatures are observed in sequence in 
the Fe~{\small XIX} line, reflecting upflows and downflows of hot plasma jets, respectively. With the 
projection effect corrected, the measured outflow speed is between $\sim$900$-$3500 km~s$^{-1}$, consistent 
with theoretical predictions of the Alfv\'{e}nic outflows in magnetic reconnection region in solar 
impulsive events. Based on theoretic models, the magnetic field strength near the reconnection region 
is estimated to be 19$-$37 Gauss. 
\end{abstract}


\section{INTRODUCTION}
It is generally accepted that the free magnetic energy stored in the coronal magnetic field is 
explosively released through magnetic reconnection \citep{pri00}, a process
in which oppositely-directed magnetic field lines break and reconnect with each other in a small area
in the corona. The residual magnetic tension in the newly-reconnected field causes the magnetic field and
plasma to be expelled from both sides of the reconnection region, forming high-speed outflows at
near-Alfv\'{e}n speed ($\sim$thousands of km~s$^{-1}$). Indirect observational evidence have been
found in past studies, e.g., the separation motion of flare ribbons formed in the lower atmosphere, cusp-shaped 
soft X-ray flare loops \citep{tsu92}, loop-top hard X-ray sources \citep{mas94}, and double hard X-ray sources 
above the flare loop top with opposite temperature gradients \citep{sui03}. 

The search for inflows and outflows is important since they are direct consequence of the reconnection
process and their measurements can yield estimates of the reconnection rate. There have been some observations
interpreted as the inflow or outflow. Apparent bidirectional motion of coronal structures around
a flare loop discovered with SOHO/EIT has been considered to be the signature of reconnection inflow 
\citep{yok01}. There has been, however, only indirect evidence of reconnection outflows, mostly based 
on morphological changes. For example, plasma blob ejections \citep{shi95} and downward plasma motions 
\citep{mck99} seen in X-ray images were thought to be a consequence of upward or downward outflows,
but the deduced outflow velocities are much lower than predicted values. 
In this paper, we present the first direct measurement of high-speed reconnection 
outflows in a flare event using combined imaging and spectroscopic observations. 

\begin{figure}
\epsscale{1.3}
\plotone{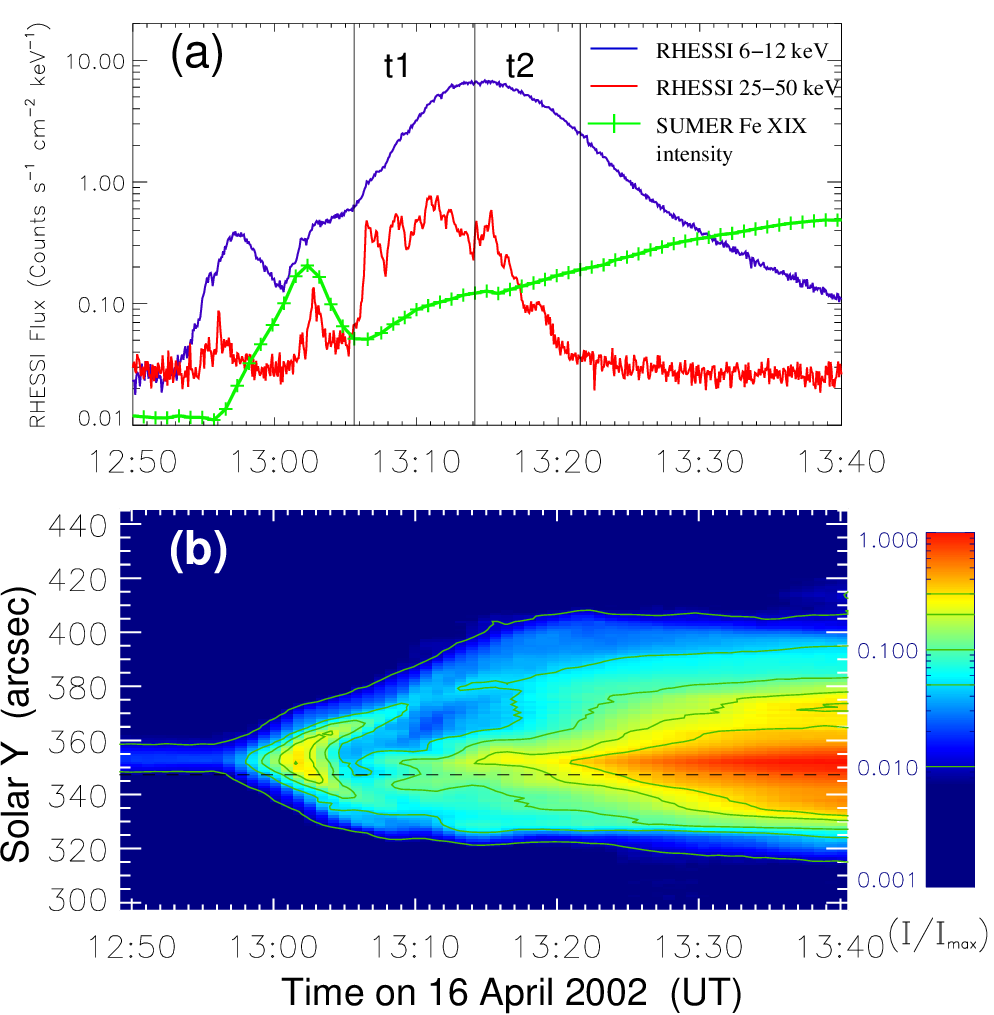}
\caption{ \label{cur}
Flux evolution of an M2.5 flare on April 16, 2002, obtained with RHESSI and SUMER.
 ({\it a}) RHESSI light curves at 6-12 (blue curve) and 25-50~keV (red curve) energy bands. The overlaid
 green curve is the time profile of the Fe~{\small XIX} line intensity at the slit position
 (dashed line in panel ({\it b})) where a high blue-shift jet (see Fig.~\ref{jet}{\it b}) was located.
 The time range marked with $t1$ is when the high blue-shift jet was seen (top two rows of
 Fig.~\ref{spc}); the time range marked $t2$ is when the high red-shift jet was seen (bottom row
 of Fig.~\ref{spc}). ({\it b}) Time series of the Fe~{\small XIX} line intensity (on a logarithmic scale)
 at a fixed slit position observed by SUMER. The overlaid contours indicate 1, 5, 10, 20, and 30\%
 of the peak intensity in the image.}
\end{figure}

\begin{figure*}
\epsscale{0.9}
\plotone{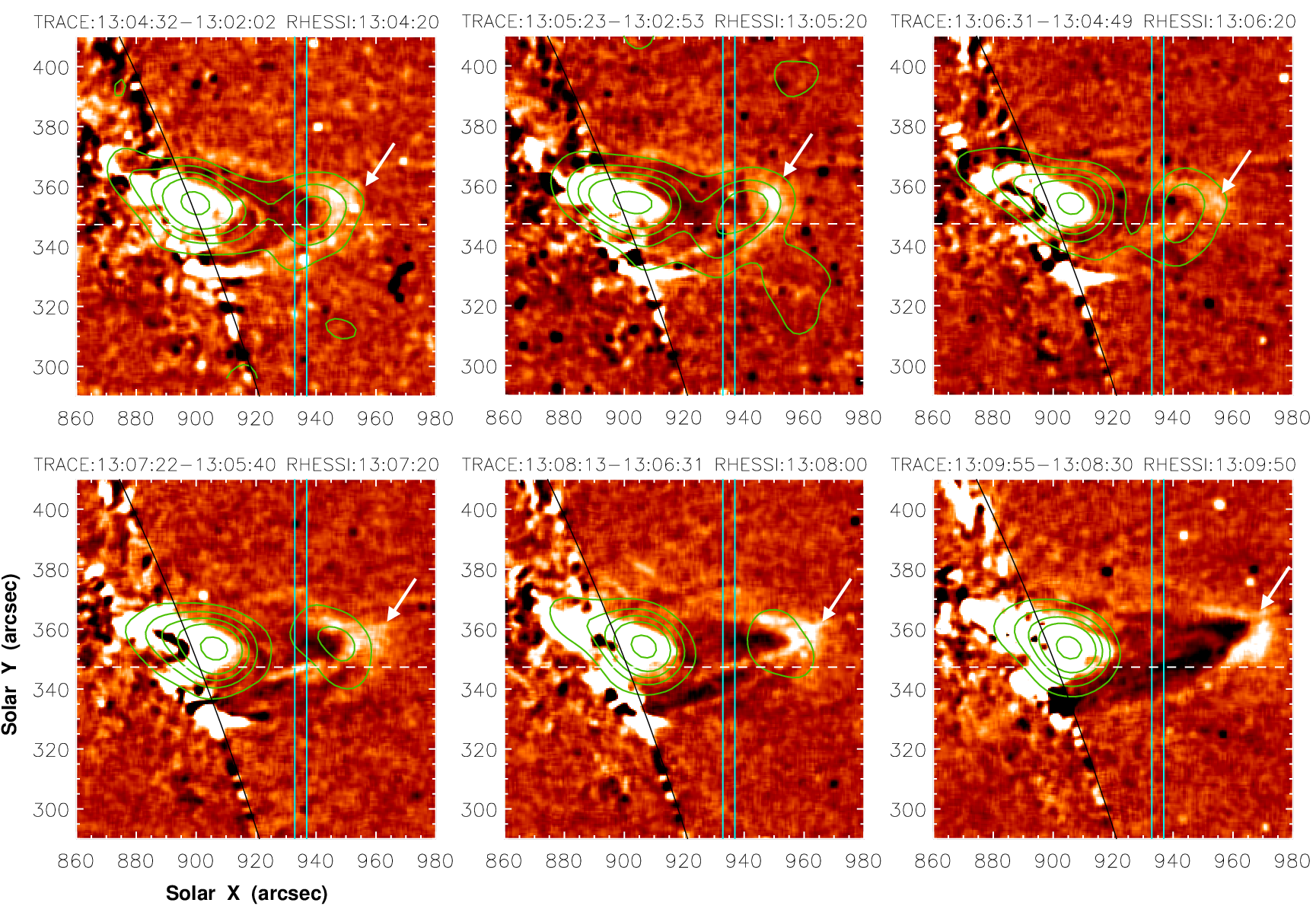}
\caption{ \label{lop}
RHESSI 6-12 keV images overlaid on the TRACE 195 \AA~ difference images during the impulsive phase of
 the flare. Evolution of a rising faint coronal loop (marked by an arrow) can been seen in TRACE
 with the dark regions representing the loop seen in the earlier map and the white regions representing
 the loop in the current map. The RHESSI contour levels are 15, 30, 45, 60, and 90\% of the peak flux
 in each image. The RHESSI images were reconstructed with the CLEAN algorithm using grid 3-9, giving a
 resolution about 7${''}$. The location of the SUMER slit is indicated by the two vertical, solid lines. 
 The horizontal dashed line indicates the position of the high blue-shift jet. An animation of this figure is available online (\textcolor{red}{http://solar.physics.montana.edu/wangtj/outflow\_mov.gif}).
}
\end{figure*}

\section{OBSERVATIONS}

The flare studied in the paper occurred very close to the northwest limb of the Sun in the active region
NOAA 9901 on 16 April 2002. It was a GOES M2.5-class flare and associated with a slow coronal mass ejection 
\citep{gof05}. The X-ray light curves in two energy bands (6$-$12 and 25$-$50~keV) obtained with the Ramaty High-Energy
Solar Spectroscopic Imager (RHESSI) are shown in Figure~\ref{cur}{\it a}. The flare started with enhancement
in soft X-rays at 12:52~UT. The impulsive phase as shown in $>$25~keV hard X-ray (HXR) started at 13:06 UT and 
lasted $\sim$11~min. The RHESSI X-ray images below 25 keV show a compact flare loop and a
separated, outward moving coronal source (Fig.~\ref{lop}, contour images). Figure~\ref{lop} also shows the
195 \AA~bandpass images obtained with the Transition Region and Coronal Explorer (TRACE), in which a bright,
compact loop and a faint, outward moving loop can be identified. \citet{gof05} reported that the faint loop
was rising with a velocity of 45$-$75 km~s$^{-1}$. The outward moving X-ray coronal source appeared to be trailing the
front of the large-scale loop seen in TRACE images (indicated by an arrow in Fig.~\ref{lop}).

\begin{figure}
\epsscale{1.2}
\plotone{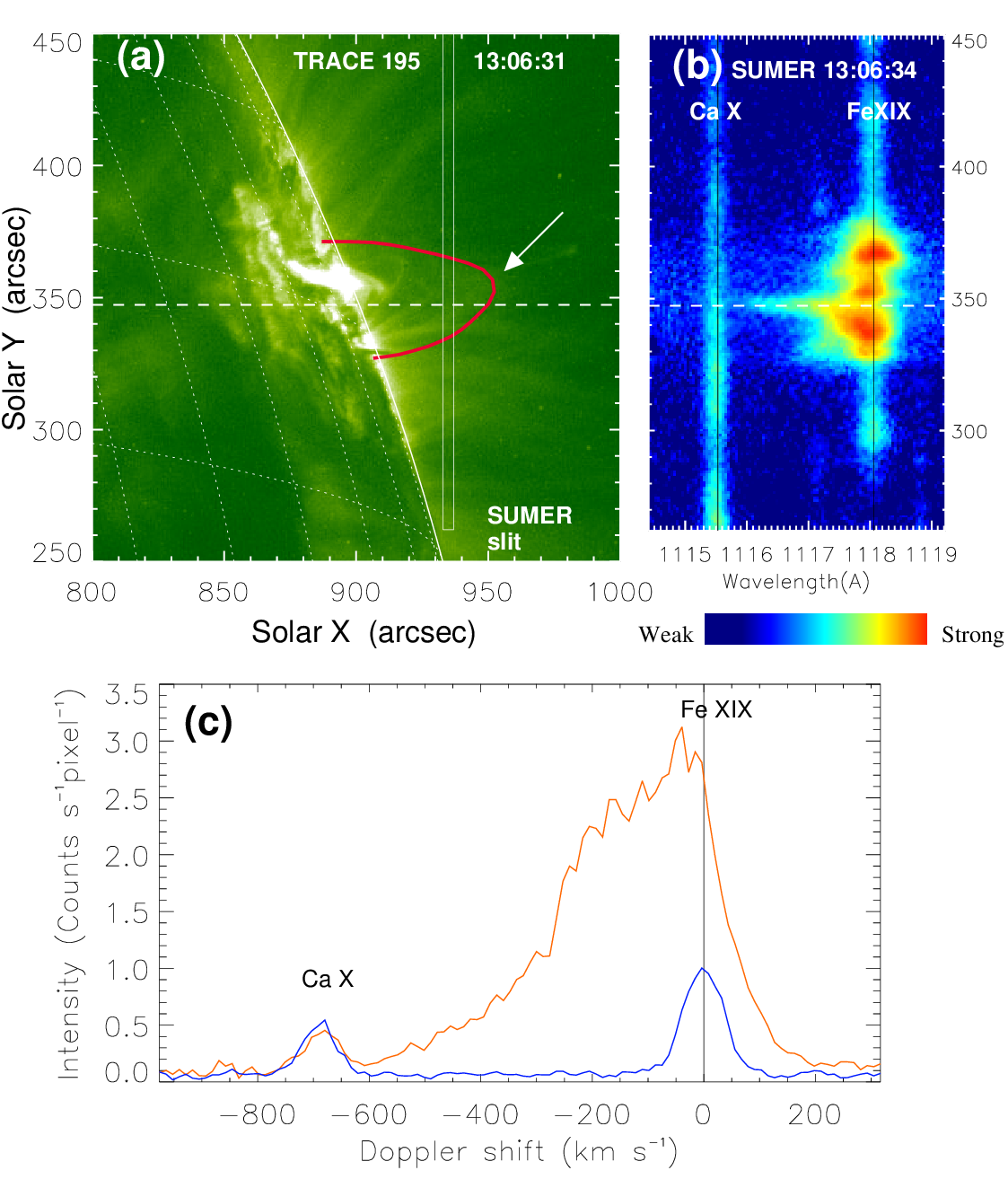}
\caption{ \label{jet}
Observations of a high-speed plasma outflow. ({\it a}) TRACE 195 \AA~ image showing the compact
 flare loop. A sketch of the faint, rising loop (red curve) as seen in the TRACE difference image (Fig.~\ref{lop})
is also indicated to show its spatial relation to the SUMER slit. The SUMER slit position is co-aligned with
 the TRACE image with an accuracy of $\sim$1$^{''}$ in the Y-direction based on common features seen
 in both the SUMER Ca~{\small X} line and TRACE intensity profiles along the slit.
({\it b}) SUMER spectra along the slit in a window containing a coronal line, Ca~{\small X} 557.7 \AA~ (2nd order)
and a hot flare line, Fe~{\small XIX} 1118.1 \AA.  ({\it c}) The spectral line profile of Fe~{\small XIX}
 (red curve) at the highly blue-shifted position (marked by the dashed line in ({\it b})), indicating a
 plasma flow with the line-of-sight component up to 600 km~s$^{-1}$.  The blue curve is the Fe~{\small XIX} line
 profile taken about a half hour before the event, showing the stationary profile.}
\end{figure}

The flare was observed throughout its duration by the Solar Ultraviolet Measurements of Emitted 
Radiation (SUMER) spectrometer on SOHO with the 300${''}\times4{''}$
entrance slit at a fixed position above the active region (Figs.~\ref{lop} and \ref{jet}{\it a}).
The spectra were recorded with a 50~s cadence in three lines: Si~{\small III} {1113.2 \AA} (0.06 MK),
Ca~{\small X} {557.7 \AA} (2nd order, 0.6 MK), and Fe~{\small XIX} {1118.1 \AA} (8 MK). 
Figure~\ref{cur}{\it b} shows the time series of the Fe~{\small XIX} line
intensity profiles along the slit. From 12:58 to 13:04~UT, the line intensity increased significantly,
appearing as a crescent-shaped structure passing the slit. The timing and location (relative to the slit) of this
structure suggest that it is the moving, large-scale (faint) loop seen by TRACE (see Fig.~\ref{lop}).
This structure is not seen in the SUMER cooler Ca~{\small X} line (Fig.~\ref{jet}{\it b}), indicating 
that the loop seen at TRACE 195 bandpass was due to emissions of hotter lines (i.e., Ca~{\small XVII} 193 \AA~ 
at 5 MK and Fe~{\small XXIV} 192 \AA~ at 20 MK) \citep{war99}. After the passage of this rising loop,
the Fe~{\small XIX} intensity decreased by about an order of magnitude (from $\sim$13:04 to 13:08 UT). Then it
gradually increased again over a larger region, seen as a cusp-shaped brightening (Fig.~\ref{cur}{\it b}), 
which indicates apparently-growing hot, dense cusp-shaped loops across the slit. This could be the evidence 
of progressive reconnection of field lines towards higher altitudes in time.

\begin{figure*}
\epsscale{1.0}
\plotone{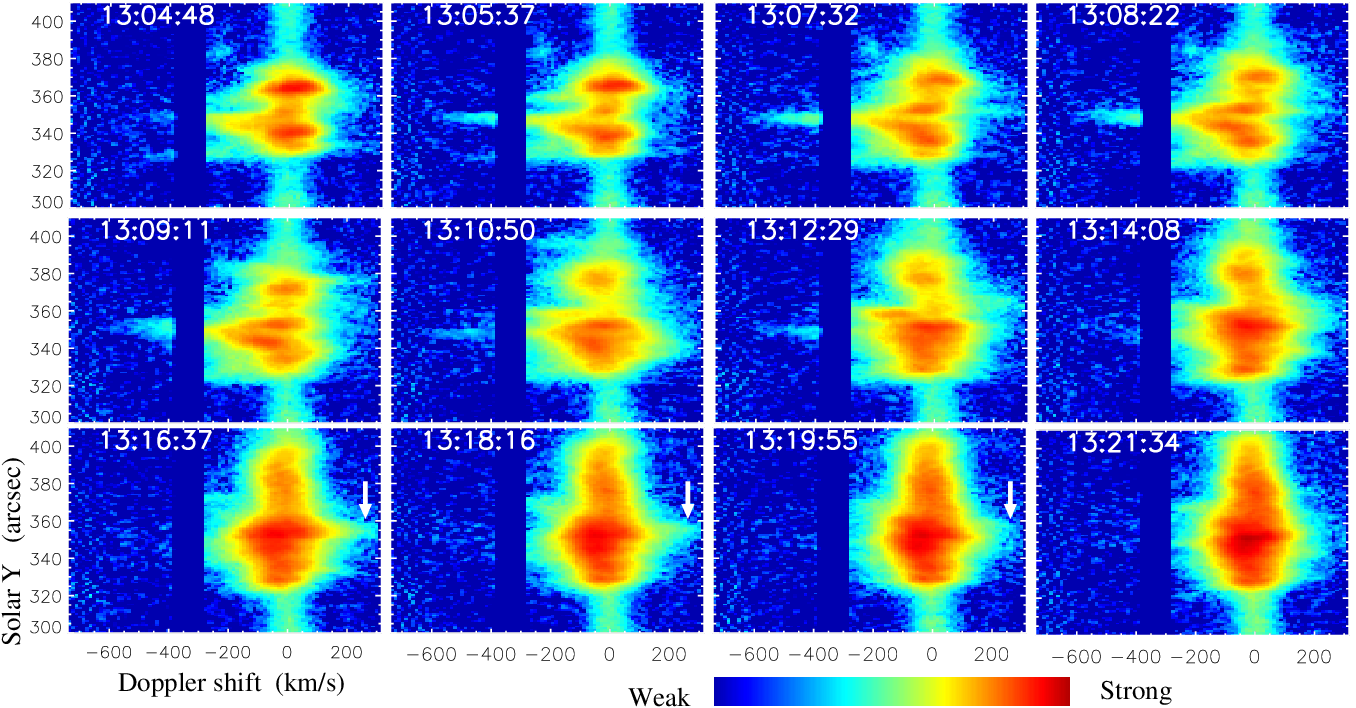}
\caption{ \label{spc}
Time series of SUMER Fe~{\small XIX} spectra showing a high blue-shift jet (top two rows) and a
 red-shift jet (marked by an arrow in bottom row), indicating a high-speed plasma upflow and downflow,
 respectively. The images are on a logarithmic scale. The continuum background and Ca~{\small X} line emissions
 have been subtracted. There is a data gap between $-$395 and $-$276 km~s$^{-1}$ in each image. An animation of this figure is available online (\textcolor{red}{http://solar.physics.montana.edu/wangtj/outflow\_mov.gif}).}
\end{figure*}

The most striking signature in SUMER observations is a plasma jet with a large blue-shift in the Fe~{\small XIX} line 
(Figs.~\ref{jet}{\it b} and \ref{spc}) lasting for $\sim$8 min during the impulsive phase of the flare. An animation for simultaneous TRACE and SUMER observations is available online\footnote{in http://solar.physics.montana.edu/wangtj/outflow\_mov.gif}.
The jet had a width of $\sim$6${''}$ along the slit, and was located at $\sim$5${''}$ south of the top of the faint, 
rising loop. This high blue-shift jet was first seen
at 13:05 UT, when the main impulsive phase of the flare started as revealed by the RHESSI 25-50~keV light
curve (Fig.~\ref{cur}{\it a}). At this time, both the top of the erupting hot loop seen by SUMER and TRACE and the X-ray
coronal source seen by RHESSI had passed the slit. Therefore, the high-speed outflow was trailing 
the coronal features moving outwards. The Fe~{\small XIX} line profile shows blue-shift component corresponding
to a line-of-sight Doppler velocity of up to 600~km~s$^{-1}$ (Fig.~\ref{jet}{\it c}). Taking into account the 
projection effect, the outflow velocity can be as high as 1800$-$3500~km~s$^{-1}$ (see discussions in 
Sect~\ref{dis}).  Figure~\ref{spc} shows the jet evolution as observed in Fe~{\small XIX} spectra. We do not see
any significant change in the line profile during the first 4 min, although the line intensity continued to increase
(Fig.~\ref{cur}{\it a}, green line). The jet became weaker after 13:10~UT, and eventually disappeared at 13:14~UT.

A high red-shift jet in the Fe~{\small XIX} line, although not as strong as the blue-shift jet, is also observed
after 13:16~UT (Fig.~\ref{spc}, bottom row). This indicates a downward-moving plasma flow passing through
the slit. The red-shift jet was at $\sim$5${''}$ north of the blue-shift jet. The maximum
Doppler velocity was 300~km~s$^{-1}$. With the projection effect corrected, we get
a downward velocity between 900 and 1800 km~s$^{-1}$. This red-shift jet disappeared after 13:21~UT when the
25-50~keV HXR emission dropped to the background level  (Fig.~\ref{cur}{\it a}).

\section{DISCUSSIONS}
\label{dis}
We note that, since the blue-shift jet seen by SUMER was located near the southern leg of the rising faint loop seen
by TRACE (Fig.~\ref{lop}, panels 3$-$5), it may be falsely interpreted as the signature of upward flows in the loop.
However, there are several arguments against such an interpretation: (1) As both the northern and southern
legs of the rising loop crossed the slit, if the blue shift is caused by the upward flow in the loop, then we would
expect to see two strong Doppler-shift jets along the slit; (2) After the southern leg of the TRACE loop
passed through the jet location, we still observed the high blue-shift jet (Fig.~\ref{lop}, bottom right panel).
(3) A plasma flow with such a high speed is not expected and has never been reported in erupting loops. 

\begin{figure*}
\epsscale{0.9}
\plotone{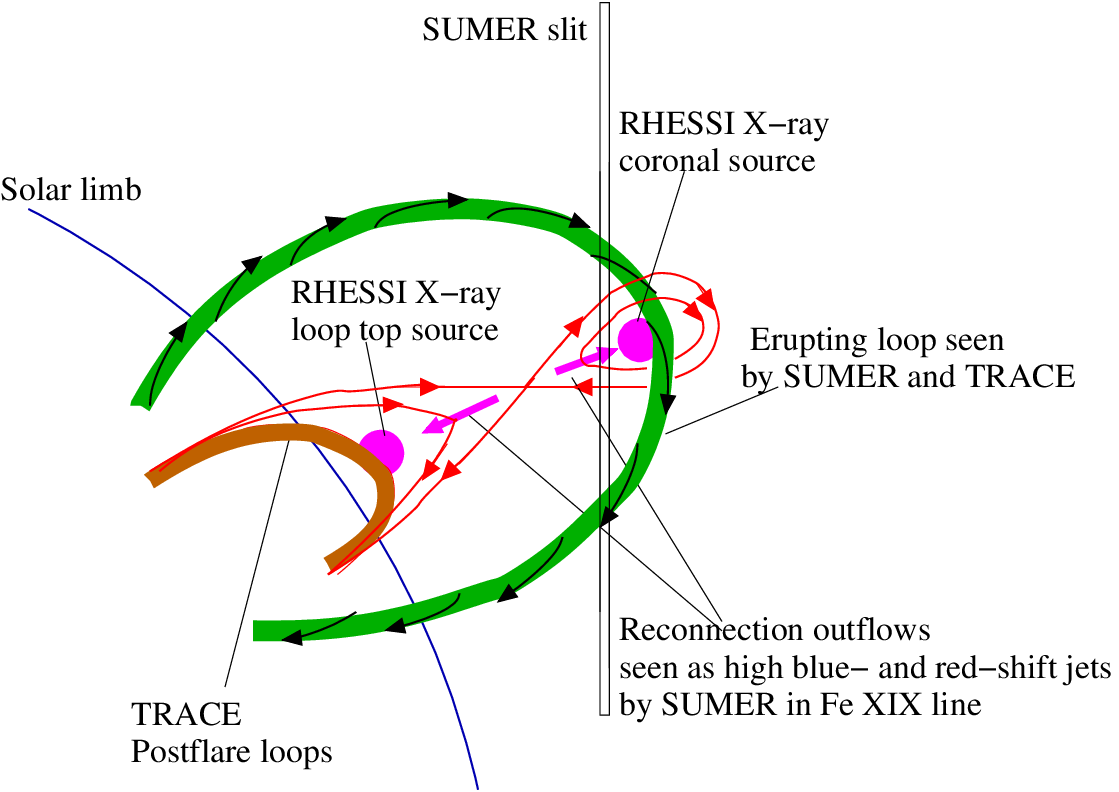}
\caption{ \label{cartn}
A schematic interpretation of the observed EUV and X-ray fea-
tures in the flare in terms of the standard flare model
}
\end{figure*}

By combining all the observations presented above, we conclude that the observed upward and downward jets are both
high-speed outflows from the magnetic reconnection site. When the upflowing jet was observed,
the reconnection site was below the slit and later when the downflowing jet was observed, the reconnection
site had moved above the slit. We interpret the essential observations in terms of the standard magnetic reconnection picture (see Fig.~\ref{cartn}): 
(1) The initial faint, rising loop observed with SUMER and TRACE is an erupting (twisted) flux rope, below which 
magnetic reconnection took place. The high temperature ($>$6 MK) of this loop suggests the presence of a pre-heating 
process which may be involved in the trigger of its eruption. (2) At the onset of the
flare impulsive phase, the high-speed upflows were expelled outward from the magnetic reconnection region.
This plasma upflow passed through the SUMER slit and produced the high blue-shift jet seen in the Fe~{\small XIX}
line. The estimated upflow speed agrees with the typical Alfv\'{e}n speed in the corona. (3) After the reconnection 
site (e.g., current sheet) passed through the slit, the observation showed the cusp-shaped, newly-reconnected field
retracting downwards and forming reconnection downflows, which were observed as the red-shift jet
by SUMER. The relatively smaller downflow velocity may imply that the downflow is slowed down by the
density increase due to chromospheric evaporation along the reconnected loops. Similar asymmetry between upflow and
downflow velocities was also found in bi-directional jets observed in explosive events in the solar 
chromosphere \citep{inn97}.  (4) After the impulsive phase, the reconnection rate
decreased significantly \citep[inferred from its temporal correlation with hard X-ray flux,][]{qiu04} and, therefore, 
no high-speed reconnection outflow can be detected with SUMER. (5) The narrow width of the observed outflows suggests 
the presence of a vertical current sheet in side view.
(6) The X-ray coronal source is closer to the reconnection region than the rising loop,
therefore, it has a higher temperature (30~MK) \citep{sui05}. Both this above-the-loop X-ray coronal source and the
lower compact X-ray source could be heated by termination shocks formed by the reconnection outflows above
and below the current sheet \citep{tsu97, mas94}. These interpretations are consistent with the results of
\citet{sui03, sui04, sui05}. Based on the RHESSI observations
that the temperature of the flare loop increased toward higher altitude and the temperature of the above-the-loop
coronal source increased toward lower altitude in the event studied here and another homologous event,
Sui et al. concluded that the magnetic reconnection occurred between the flare loop and the above-the-loop
coronal source. 

\begin{figure}
\epsscale{1.2}
\plotone{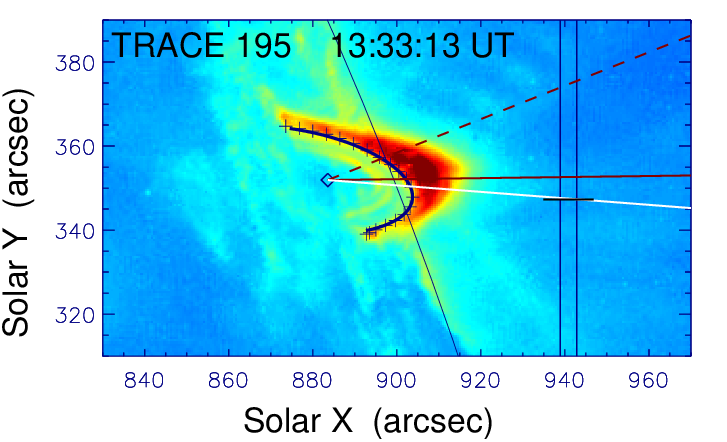}
\caption{ \label{ang}
 Postflare loops seen in TRACE 195\AA~image at 13:33 UT. The inner side of the loop (outlined with plus signs) is
fitted with a circular arc (black curve) in a 3-D geometry. 
The radial direction (dashed line) from the midpoint of the loop baseline
has an angle to the line-of-sight (LOS) of 80$^{\circ}$. The line linking
the midpoint of the two footpoints to the loop apex (dark solid line) is indicated to have
an orientation towards ($\sim$5$^{''}$ above) the high-shift jet seen by SUMER. The direction of this line
has an angle to the LOS of 70$^{\circ}$. The white line representing a direction in the loop plane and pointing
just towards the jet position has an angle to the LOS of 80$^{\circ}$. Therefore, we assume the outflows in the
direction which has an angle to the LOS between 70$^{\circ}$ and 80$^{\circ}$ in the text. 
}
\end{figure}

Now we estimate the true jet speed based on Figure~\ref{ang}. If we assume the outflow is radial from the active region, 
then it would be at $\sim$80$^{\circ}$ relative to the line-of-sight. If we fit the inner side of the TRACE compact 
flare loops by a circular arc in a 3-dimensional geometry \citep{asc02} 
and assume that the outflow is in the direction of the line linking the midpoint of the two loop feet and
the loop apex, the angle is $\sim$70$^{\circ}$. Thus, the maximum blue shift of 600 km~s$^{-1}$ gives an outflow
velocity as high as 1800$-$3500~km~s$^{-1}$ taking into account the projection effect. 
Since the reconnection outflows had a velocity close to the Alfv\'{e}n speed
[$V_A\simeq3000(B/20~{\rm G}) (n_e/2\times10^8~{\rm cm^{-3}})^{-1/2}$ km~s$^{-1}$] (where $B$ is the magnetic
field and $n_e$ is the electron density), we can estimate the magnetic field strength near the reconnecting
region. Given $n_e\sim5\times10^8$ cm$^{-3}$ as the average active region coronal density at a height of
$\sim$40 Mm (SUMER slit height) \citep{del03}, with $V_A$ of 1800$-$3500 km~s$^{-1}$,
we get $B\simeq$19$-$37 Gauss, which agrees well with the mean magnetic field strength of coronal loops at
similar heights measured from observations of coronal loop oscillations \citep{nak01, asc02, wan07}.
The measurements of outflow speed and magnetic field strength near the reconnection region are
essential for our understanding of the plasma heating and particle acceleration in flares.

High temperature flows of $\sim$1000 km~s$^{-1}$ have been observed with SUMER in a couple of other flares 
\citep{inn01, inn03}. However, since observations of those events provide no information of 
the location of the reconnection region, it is difficult to pinpoint the origin of these flows, 
or directly relate these flows to the energy release region. For example, \citet{inn03} observed blue shifts 
of 800$-$1000 km~s$^{-1}$ in the Fe~{\small XXI} line from the boundary along the tail of dark downflows.
The physical nature of dark downflows is not fully understood. \citet{mck99} interpreted the dark 
downflows as signatures of reconnection outflows, though the speeds are slower than typically expected and 
high Doppler shifts at the tail boundary are hardly explained in their picture. \citet{inn03}
suggested another way that these shifts may be due to a fast wind generated behind the downflows.
 
The combined observations of SUMER, RHESSI, and TRACE of a very well observed event studied in this paper
allow us to determine unambiguously the spatial relationship between the high Doppler-shift 
flows and the reconnection region, and thus provide direct evidence of high-speed magnetic reconnection 
outflows in the current sheet in the corona. The observations lend strong support to the magnetic 
reconnection theory and the bipolar reconnection model of solar eruptive events. To better understand  
magnetic reconnection on the Sun, it requires a further analysis based on 3D models
which have shown many different properties from the 2D CSHKP model \citep[e.g.,][]{arc05}.

\acknowledgments
We thank D. E. Innes, B. R. Dennis and G. D. Holman for their valuable discussions and suggestions.
We also thank Max-Planck-Institut f\"{u}r Sonnensystemforschung for providing the SUMER data.
T.W. and J.Q.'s work are supported by NASA grant NNG06GA37G and NASA grant NAS5-38099. L.S.'s work is supported 
by NASA grant 370-16-20-16 and the RHESSI project.


\begin{thebibliography}{}
\bibitem[Archontis et al. (2005)]{arc05} Archontis, V., Moreno-Insertis, F., Galsgaard, K., \& Hood, A. W.
   2005, \apj, 635, 1299
\bibitem[Aschwanden et al. (2002)]{asc02} Aschwanden, M. J., et al.  2002, \solphys, 206, 99
\bibitem[Del Zanna \& Mason (2003)]{del03} Del Zanna, G., \&  Mason, H. E., \aap, 406, 1089
\bibitem[Goff et al.(2005)]{gof05} Goff, C. P.,  et al. 2005, \aap,  434, 761
\bibitem[Innes et al. (1997)]{inn97} Innes, D. E., et al. 1997, Nature, 386, 811
\bibitem[Innes et al. (2001)]{inn01} Innes, D. E., et al. 2001, \apjl, 549, L249
\bibitem[Innes, McKenzie, \& Wang (2003)]{inn03}  Innes, D. E., McKenzie, D. E., \&  Wang, T. J. 2003,
 \solphys, 217, 267
\bibitem[Masuda et al. (1994)]{mas94} Masuda, S., et al. 1994, Nature, 371, 495
\bibitem[McKenzie \& Hudson (1999)]{mck99} McKenzie, D. E., \& Hudson, H. S. 1999, \apjl, 519, L93
\bibitem[Nakariakov \& Ofman (2001)]{nak01} Nakariakov, V. M., \& Ofman, L. 2001, \aap, 372, L53
\bibitem[Priest \& Forbes (2000)]{pri00} Priest, E. R., \& Forbes, T. 2000, Magnetic Reconnection - MHD Theory and 
    Applications (Cambridge Univ. Press, Cambridge, 2000).
\bibitem[Qiu et al. (2004)]{qiu04} Qiu, J., et al., 2004, \apj, 604, 900
\bibitem[Shibata et al. (1995)]{shi95} Shibata, K., et al. 1995, \apjl, 451, L83
\bibitem[Sui (2004)]{sui05} Sui, L. 2004, Ph.D. thesis (Catholic Univ. America)
\bibitem[Sui \& Holman (2003)]{sui03} Sui, L., \& Holman, G. D. 2003, \apjl, 596, L251
\bibitem[Sui, Holman, \& Dennis (2004)]{sui04} Sui, L., Holman, G. D., \& Dennis, B. R. 2004, \apj, 612, 546
\bibitem[Tsuneta et al. (1992)]{tsu92} Tsuneta, S., et al. 1992, PASJ, 44, L63 
\bibitem[Tsuneta (1997)]{tsu97} Tsuneta, S. 1997, \apj, 483, 507
\bibitem[Wang, Innes, \& Qiu (2007)]{wan07} Wang, T. J., Innes, D. E., \& Qiu, J. 2007, \apj, 656, 598
\bibitem[Warren et al. (1999)]{war99} Warren, H. P., et al. 1999, \apjl, 527, L121
\bibitem[Yokoyama et al. (2001)]{yok01} Yokoyama, T., et al. 2001, \apjl, 546, L69

\end{thebibliography}
\end{document}